\documentstyle[aps,prl,twocolumn]{revtex}

\def\be{\begin{equation}}
\def\ee{\end{equation}}
\def\bea{\begin{eqnarray}}
\def\eea{\end{eqnarray}}
\def\sp{{\mathcal C}}
\def\tr{{\rm tr}}
\def\1/2{\frac{1}{2}}

\newcommand{\la}{\langle}
\newcommand{\ra}{\rangle}

\begin{document}

\draft
\wideabs{

\title{Distillability, Bell inequalities and multiparticle bound entanglement}

\author{A. Ac\'\i n}

\address{Departament d'Estructura i Constituents de la
Mat\`eria, Universitat de Barcelona, Diagonal 647, E-08028 Barcelona, Spain.\\
e-mail: acin@ecm.ub.es
}

\date{\today}

\maketitle


\begin{abstract}

We study the relation between violation of Bell inequalities and distillability properties of quantum states. Recently, D\"ur \cite{dur} has shown that there are some multiparticle bound entangled states, non-separable and non-distillable, that violate a Bell inequality. We prove that for all the states violating this inequality there exist at least one splitting of the parties into two groups such that some pure-state entanglement can be distilled, obtaining a connection between Bell inequalities and bipartite distillable entanglement.

\end{abstract}

\pacs{PACS Nos. 03.67.-a, 03.65.Bz}}



Non-locality or entanglement is one of the most striking properties of Quantum Mechanics. From a fundamental point of view, its importance is related to the fact that no local-realistic theory (LR) is able to reproduce the correlations observed in some entangled states of composite systems \cite{sep}. Moreover, recent results in quantum information theory show that it is also a useful resource for information transmission and processing. In this letter we find a connection between these two features of entangled states.

In their seminal work of 1935 \cite{EPR}, Einstein, Podolsky and Rosen, pointed out a conflict between the correlations appearing in some quantum states of composite systems and LR theories. Later, Bell \cite{Bell} derived some conditions, known as Bell inequalities, that are satisfied by any LR theory, but that are violated for some quantum states. The  experimental check of the violation of these inequalities \cite{ADR} confirmed that it is not possible to build a local hidden variable model (LHV) reproducing all the correlations observed for quantum states of composite systems (see also \cite{tests}). Thus, Quantum Mechanics is said to be non-local. More recently, Gisin \cite{Gisin} proved that all entangled pure states of bipartite systems violate the CHSH inequality \cite{CHSH}, and this result was also extended to multipartite entangled pure states by Popescu and Rohrlich \cite{PR}. For pure states, a Bell inequality is violated if and only if the state is not separable.

For mixed states the picture is not as clear. Indeed, Werner \cite{Werner} constructed a local hidden variables model reproducing all the correlations under Von Neumann measurements observed in some entangled mixed states, the so-called Werner states (see \cite{Barrett} for the extension of the LHV model to more general measurements). Later, Popescu \cite{Popescu} proved that some of these states can violate the CHSH inequality after a sequence a local measurements, which are able to reveal a hidden non-locality of the state (similar results were found in \cite{Gisin2}). Thus, there is a lack of a complete classification of mixed states according to their non-local properties.

On the other hand, entanglement has been also proved to be a useful resource for information processing. Most of the new quantum information applications, such as for instance dense-coding \cite{sdense} or teleportation \cite{telep}, use maximally entangled pure states for achieving some results that have not analogue in Classical Information Theory. In practical situations, and because of decoherence, we do not deal with entangled pure states but with entangled mixed states, from which we have to {\em distill}, using only local operations and classical communication (LOCC), some amount of pure-state entanglement. Indeed it is known that all pure states with non-zero entanglement can be transformed by LOCC, with some probability, into maximally entangled states, but for density matrices  we do not have any criterion for knowing whether a state is distillable or not (see \cite{primer} and references therein). It was shown by the Horodecki \cite{horo} that there exist some mixed states, known as bound entangled states, which cannot be distilled, i.e. they are not useful for any quantum information task, in spite of being entangled.

It would be very interesting to find some relations between non-local features of mixed states in terms of violation of Bell inequalities, and distillability properties. Intuitively, one is tempted to say that if a state violates a Bell inequality and its correlations cannot be described by a LHV model (they are intrinsically quantum), they are useful for quantum information processing, i.e. the state is distillable. Following with this intuition, bound entangled states are conjectured to be density matrices that can be described by a LHV model, despite being non-separable  \cite{Peres}. Some results in this direction have been obtained in \cite{Terhal,WW,WW2,KZG}.

However, D\"ur \cite{dur} has recently shown that there are some multipartite bound entangled states, non-separable and non-distillable, that can violate some Bell inequality. From his result it follows that the violation of a Bell inequality does not imply distillability, and that some bound entangled states contradict local realism. 

In this letter, starting from the same Bell inequality as in \cite{dur}, we demostrate that it is still possible to find a link between its violation and distillability. In fact, as it is proved below, for all the states violating it, there is at least one bipartite splitting of the system such that the state becomes distillable. Thus, either all these states are distillable or they have some bound entanglement that can be activated \cite{DC} after joining some of the parties.


The Bell inequality D\"ur considered is the Mermin-Klyshko inequality \cite{Mermin,BK}, which generalizes the CHSH inequality for $N$-qubit systems, when each party, $i$, measure the two dichotomic observables $\sigma_{\hat n_i}=\hat n_i\cdot\sigma$ and $\sigma_{\hat n'_i}=\hat n'_i\cdot\sigma$. It reads (see also \cite{WW,WW2,GBP,ZB,SG})
\be
\label{mermin}
\la B_N\ra\leq 1 ,
\ee
where $B_N$ is the Bell operator defined recursively as
\be
B_i=\1/2B_{i-1}\otimes(\sigma_{\hat n_i}+\sigma_{\hat n'_i})+\1/2B'_{i-1}\otimes(\sigma_{\hat n_i}-\sigma_{\hat n'_i}),
\ee
$B'_i$ is obtained from $B_i$ by exchanging all the $\hat n_i$ and $\hat n'_i$ and $B_1=\sigma_{\hat n_1}$. The chosen measurement directions are $\sigma_{\hat n_i}=\sigma_x$ and $\sigma_{\hat n'_i}=\sigma_y$, $\forall i$, and this gives
\be
\label{durin}
B_N=2^{\frac{N-1}{2}}(e^{i\beta_N}|1^{\otimes N}\ra\la 0^{\otimes N}|+e^{-i\beta_N}|0^{\otimes N}\ra\la 1^{\otimes N}|) ,
\ee
with $\beta_N=\pi/4(N-1)$. Notice that these are the values that give the maximal violation, allowed by Quantum Mechanics, among the set of inequalities based on two dichotomic observables per site \cite{WW,SG}. 

The state studied in \cite{dur} is 
\be
\label{durst}
\rho^{BE}_N=\frac{1}{N+1}\left(|\Psi\ra\la\Psi|+\1/2\sum_{i=1}^N(P_i+\bar P_i)\right) ,
\ee
where $|\Psi\ra$ is a $N$-party Greenberger-Horne-Zeilinger (GHZ) state \cite{GHZ},
\be
|\Psi\ra=\frac{1}{\sqrt 2}(|0^{\otimes N}\ra+e^{i\alpha_N}|1^{\otimes N}\ra) ,
\ee
$\alpha_N$ being a phase factor. By $P_i$ it is denoted the projector onto the state $|\phi_i\ra$, which is a product state equal to $|1\ra$ for party $i$ and $|0\ra$ for the rest, and $\bar P_i$ is obtained from $P_i$ permuting zeros and ones. It is proved that $\rho^{BE}_N$ is bound entangled for $N\geq 4$, since all the operators $(\rho^{BE}_N)^{T_i}$ are positive, where $T_i$ is the partial transposition with respect to party $i$, and it violates (\ref{durin}) for $N\geq 8$, with $\alpha_N=\beta_N$.

Let us come back to the Bell operator (\ref{durin}). The phase factor can be absorbed after local phase redefinition and $B_N$ can be written as
\be
\label{ineq}
B_N=2^{\frac{N-1}{2}}(|\Psi_0^+\ra\la\Psi_0^+|-|\Psi_0^-\ra\la\Psi_0^-|) ,
\ee
where $|\Psi_0^\pm\ra$ are the GHZ states
\be
\label{GHZ}
|\Psi_0^\pm\ra=\frac{1}{\sqrt 2}(|0^{\otimes N}\ra\pm|1^{\otimes N}\ra) .
\ee
Now we can prove the main result of the letter.

{\em Lemma:} For all the $N$-qubit states $\rho$ violating the Bell inequality (\ref{durin}), i.e. $\tr(B_N\rho)>1$, there exists a bipartite splitting of the composite system such that the state is distillable (in the sense of \cite{DC}).

{\em Proof:} Consider a state with $\tr(B_N\rho)>1$ and apply to it the depolarization protocol described in \cite{DC2}. The state is transformed into one of the members of the family of $N$-qubit states $\rho_N$ (see \cite{DC2} for details)
\bea
\label{family}
\rho_N=&&\sum_{\sigma=\pm}\lambda_0^\sigma|\Psi_0^\sigma\ra\la\Psi_0^\sigma| \nonumber\\
&&+\sum_{j=1}^{2^{N-1}-1}\lambda_j(|\Psi_j^+\ra\la\Psi_j^+|+|\Psi_j^-\ra\la\Psi_j^-|) ,
\eea
where 
\be
\label{GHZj}
|\Psi_j^\pm\ra\equiv\frac{1}{\sqrt 2}(|j\ra|0\ra\pm|2^{N-1}-j-1\ra|1\ra) ,
\ee
and $j=j_1j_2\ldots j_{N-1}$ is understood in binary notation. The values of the positive coefficients appearing in (\ref{family}) are kept unchanged during the depolarization protocol, i.e. $\lambda_0^\pm=\la\Psi_0^\pm|\rho_N|\Psi_0^\pm\ra=\la\Psi_0^\pm|\rho|\Psi_0^\pm\ra$ and $2\lambda_j=\la\Psi_j^+|\rho_N|\Psi_j^+\ra+\la\Psi_j^-|\rho_N|\Psi_j^-\ra=\la\Psi_j^+|\rho|\Psi_j^+\ra+\la\Psi_j^-|\rho|\Psi_j^-\ra$. Since $\rho$ violates (\ref{durin}) we have
\be
\label{condition}
\Delta\equiv \lambda_0^+-\lambda_0^->\frac{1}{2^{\frac{N-1}{2}}} ,
\ee
while from normalization
\be
\label{norm}
\lambda_0^++\lambda_0^-+2\sum_j\lambda_j=1 .
\ee

For the family of states $\rho_N$ there exists a nice correspondence between their distillability (and separability) properties and the coefficients $\{\lambda_0^\pm,\lambda_j\}$. Consider the set ${\mathcal P}$ of all possible bipartite splittings of the $N$ particles into two groups. Any element of this set can be denoted by $P_j$, where $j=j_1j_2\ldots j_{N-1}$ is a string of $N-1$ bits with $j_i=0$ if party $i$ belongs to the same set as the last party. For example, for three qubits the splittings $(13)-(2)$, $(23)-(1)$ and $(3)-(12)$ are given by $P_{01}$, $P_{10}$ and $P_{11}$. Note that the string $j$ cannot be zero and there are $2^{N-1}-1$ of such splittings (as $\lambda_j$ coefficients). It was proved in \cite{DC} that the state $\rho_N$ is bipartite distillable for the splitting $P_j$ if and only if
\be
\label{DCcond}
2\lambda_j<\Delta .
\ee

Now come back to the state $\rho_N$ obtained from a state $\rho$ violating the Mermin inequality (\ref{durin}), and consider the case in which there is no bipartite splitting such that $\rho_N$ is distillable (or the bound entanglement can not be activated). This means that $2\lambda_j\geq\Delta,\,\forall j$, and summing over $j$ and using (\ref{condition}) it follows that
\be
2\sum_j\lambda_j\geq (2^{N-1}-1)\Delta>\frac{2^{N-1}-1}{2^{\frac{N-1}{2}}} .
\ee
Note that the RHS of this expression is greater than one for $N>2$ (it is well known that all inseparable two-qubit states are distillable \cite{horo2}), and this is not possible because of the normalization condition (\ref{norm}). Then, we conclude that there exists at least one bipartite splitting such that $\rho_N$, and therefore the original state $\rho$ violating (\ref{durin}), is distillable. This ends the proof. $\Box$

Let us mention here that in the derivation of this lemma we have weakened the concept of distillability in such a way that it is not still excluded that the violation of a Bell inequality implies the distillability of the state (indeed this is the case for the inequality analyzed). Moreover, our result sheds some light on the fact that there are some multiparticle bound entangled states that cannot be described by a local hidden variables model, since the state may become non-local (distillable) after joining some of the parties. Note that the Mermin inequality (\ref{durin}) can be interpreted as a detector, or witness, of bipartite distillable entanglement, and the corresponding distillation protocol is given with the proof of the lemma (actually, it is the same as in \cite{DC}). Thus, any state violating this inequality, perhaps after a sequence of local operations and classical communication, is bipartite distillable.

In this letter we have shown a connection between distillability and Bell inequalities for $N$-qubit states. As far as we know, it is one of the first non-trivial results linking these two features for systems of dimension greater than two qubits. There are still many open questions concerning the relations between the different notions of non-locality, that is, Bell inequalities, partial transposition \cite{PPT} and distillability, and it would be interesting to find similar results for states of two qudits, $\sp^d\otimes\sp^d$, or for other Bell inequalities in $N$-qubit systems.

\bigskip

The author thanks Enric Jan\'e, Llu\'\i s Masanes and Guifr\'e Vidal for useful comments. Financial support from Spanish MEC (AP-98) and ESF-QIT is also acknowledged.


\begin{references}
\bibitem{sep}
Take a composite system of $N$ particles, $\sp^{d_1}\otimes\sp^{d_2}\otimes\cdots\otimes\sp^{d_N}$, where $d_i$ is the dimension of the local space of party $i$. A pure state, $|\Psi\ra$, belonging to the global space is said to be separable when it can be written as a tensor product of pure states in each subsystem, i.e. $|\Psi\ra=|\psi_1\ra\otimes|\psi_2\ra\otimes\cdots\otimes|\psi_N\ra$ with $|\psi_i\ra\in\sp^{d_i},\,\forall i$. This definition is easily extended to mixed states, and a density matrix $\rho$ is separable when it can be expressed as a convex sum of projectors onto separable pure states, i.e. $\rho=\sum_j p_j |\Psi^s_j\ra\la\Psi^s_j|$, where $|\Psi^s_j\ra$ are separable $\forall j$. States that are not separable are called entangled.
\end{references}
\end{document}